\pdfoutput=1
\documentclass[11pt,a4paper]{article}
\usepackage{jheppub}
\usepackage{epstopdf}

\newcommand {\be} {\begin{equation}} 
\newcommand {\ba}{\begin{eqnarray}} 
\newcommand {\ee} {\end{equation}} 
\newcommand{\ea} {\end{eqnarray}}

\begin{document}
	\title{Superconducting vortex in a deconstructed  holographic model}
	\author{Zainul Abidin,} 
	\author{Herry J. Kwee}
	\author{and Jong Anly Tan}
	\affiliation{STKIP Surya (Surya College of Education),\\
		Tangerang, Indonesia} 
	\abstract{A  deconstructed holographic model for an $s$-wave superconductor in $2+1$ dimensions has been proposed in the literature.  Here, we consider solutions in the presence of perpendicular magnetic fields, where vortex formed. We calculate the expectation value of the Cooper pair condensate as a function of radial distance from the center of the vortex for different numbers of lattice points.}
	\emailAdd{zainul.abidin@stkipsurya.ac.id}
	\emailAdd{herry.kwee@stkipsurya.ac.id}
	\emailAdd{jongtan@stkipsurya.ac.id}
	\keywords{AdS/CFT, holography, superconductor, deconstruction}
	
	\maketitle
	\section{Introduction}
	 High-temperature superconductors were first discovered by Bednorz and Muller in 1986 in cuprate perovskite materials~\cite{Bednorz:1986tc}. After that, many more superconductors with relatively high critical temperature were discovered~\cite{Wu:1987te}. Physical mechanism behind these non-conventional superconductors remains unclear. Unlike conventional superconductors, they can't be explained by the standard BCS theory~\cite{Bardeen:1957kj}.
	 
	 Apart from the study of superconductors, in the study of string theory, Maldacena~\cite{Maldacena:1997re} conjectured a remarkable connection between a strongly coupled field theories to a gravitational theory in higher spacetime dimensions. This connection called AdS/CFT correspondence, because it relates a conformal field theory (CFT) in $d$-dimensions to a theory with gravity in Anti-de Sitter space in $d+1$-dimensions. It is also called holographic because operators in the field theory correspond to the boundary of the fields in the higher dimensional gravity theory.
	
	Application of the principle of holography to a superconductor is initiated by ~\cite{Hartnoll:2008vx, Hartnoll:2008kx}. They successfully demonstrated how a simple gravity system holographically corresponds to a material that has properties of superconductors. In the model they put forward, there is a critical temperature, $T_c$. When the material temperature is below this $T_c$, condensate is formed through second order phase transition. This condensate consists of a pair of quasiparticle.
	
	Many aspects of superconductivity have been studied using the holographic principles. The gravity dual of a superconducting system in the presence of perpendicular magnetic fields is proposed in~\cite{Hartnoll:2007ai}. It was shown how the Hall effect appears in the holographic model. Thermal and electrical conductivity has been considered in~\cite{Hartnoll:2008hs}. Modelling for the $p$-wave superconducting material is constructed in~\cite{Gubser:2008wv}. 
	
	Holographic superconductor with vortex configuration is studied in~\cite{Montull:2009fe, Albash:2009iq}. In the paper~\cite{Montull:2009fe}, the free energy as a function of external magnetic field is calculated. It was shown that in a certain range of the magnetic field, vortex configuration is energetically favorable.
	
	In the AdS/CFT correspondence, the dual extra-dimensional model is generally non-renormalizable. In order to make these models well-defined, a procedure was developed by dimensionally ``deconstructing'' the extra-dimensional theory into a renormalizable field theory defined in lower dimensions. This deconstructed model, at a long distance, should reduce to the extra-dimensional model latticized in extra-dimension.  A dimensionally deconstructed model of an $s$-wave superconductor is discussed in~\cite{Albrecht:2012ek}.    
	
	In this paper, we consider a model of a two-dimensional sheet of superconductor in the presence of a perpendicular magnetic field. As in~\cite{Albrecht:2012ek}, we consider a model of $s$-wave superconductor with deconstructed extra-dimension. We calculate the expectation value of the condensate as a function of radial distance from the center of the vortex for $N=5$, $N=10$ and $N=100$ lattice points in the extra-dimensions. 
	
	\section{Continuum Model}
	A model of superconductor in $2+1$ dimensions is holographically dual with a model of gravity in $3+1$ Anti-de Sitter space. The metric of the 3-dimensional AdS with Schwarzschild black hole is given by
	\begin{align}
		ds^2=\frac{1}{z^2}\left( f(z)dt^2-\frac{1}{f(z)}dz^2-(dx^2+dy^2)\right)\,,
	\end{align}
	where 
	\begin{align}
		f(z)=\frac{1}{L^2}\left(1-\frac{z^3}{z_h^3}\right)\,.
	\end{align}	
	The constant $L$ is the AdS radius which we will set to $1$. The coordinates are $(t,z,x,y)$ where $t$ is time, $z$ is the extra-dimension, $x$ and $y$  are  the remaining $2$-spatial dimensions. The extra-dimension runs from $z=\varepsilon$ with $\varepsilon\rightarrow 0$, the ultraviolet (UV) boundary, to $z_h$ the black hole horizon. The temperature in the strongly coupled theory can be identified as the Hawking temperature of the black hole,
	\begin{align}
		T=\frac{3}{4\pi z_h^2}
	\end{align}

	The model consists of a Maxwell field coupled to a complex scalar field. The gauge field corresponds to electromagnetism and the complex scalar field corresponds to Cooper pairs. We will consider only the probe limit, where the back-reaction of the fields on the geometry is neglected. This can be achieved by setting the gauge field coupling to be very large, which we will set to one, such that the matter sector decouples from the gravity sector.  The matter part of the action can be written as
	\begin{align}
		S=\int d^4x \sqrt{-G}\left[-\frac{1}{4}F_{MN}F^{MN}+|(\partial_M-iA_M)\psi|^2-m^2|\psi|^2\right] \label{matteraction}
	\end{align}
	where contraction of the indices $M=0,1,2,3$ is through the 4D metric $G_{MN}$. For the spatial dimension, we will use index $a,b=x,y$. The field strength of the U$(1)$ gauge field is $F_{MN}=\partial_MA_N-\partial_NA_M$. 
	
	There are two possible values for the scalar mass, $m^2=0,-2$. We will use $m^2=-2$, which corresponds to a Cooper pair operator of dimension $2$.	We consider solutions for the fields given by $\psi=\psi(z)$, $A_0=A_0(z)$ and we set $A_a=A_z=0$. We obtain coupled equation of motions 
	\begin{align}
		 \partial^2_z\psi+\left(\frac{f'}{f}-\frac{2}{z}\right)\partial_z\psi +\frac{A_0^2}{f^2}\psi+\frac{2}{z^2 f}\psi&=0\,,\nonumber\\
		 \partial_z^2A_0-\frac{2\psi^2}{z^2f}A_0&=0\,,
	\end{align} 
	where $f'\equiv\partial_z f$. Near the UV boundary, the fields behave as follow
	\begin{align}
		\psi=&\psi^{(1)} z+\psi^{(2)}z^2+\ldots\,,\\
		A_0=&\mu-\rho z+\ldots\,,
	\end{align}
	where according to AdS/CFT dictionary $\mu$ can be identified as the chemical potential and $\rho$ as the charge density. The coefficient $\psi^{(1)}$ can be identified as the source of the Cooper pair operator, and $\psi^{(2)}$ as its vacuum expectation value. Since we want condensation to occurs dynamically without source, we will set $\psi^{(1)}=0$, which equivalent to Neumann boundary condition, $\partial_z\psi(\varepsilon)=0$. 
	
	For regular solutions, the fields should satisfy the following equations near the horizon,
	\begin{align}
		\partial_z \psi(z_h)=&\frac{2\psi(z_h)}{3z_h}\,,\nonumber\\
		A_0(z_h)=&0\,,\nonumber\\
		\partial_z^2\psi(z_h)=&-\frac{4}{9z_h^2}\psi(z_h)-\frac{z_h^2}{18}(\partial_zA_0)^2\psi(z_h)\,,\nonumber\\
		\partial_z^2A_0(z_h)=&-\frac{2\psi^2(z_h)}{3z_h}\partial_zA_0(z_h)\,.
	\end{align}
	Note that there are two independent parameters near horizon $\psi(z_h)$ and $\partial_zA_0(z_h)$. To solve the equation of motions, one can use finite difference method, starting from the horizon then shoot toward the UV boundary. The value of  $\psi(z_h)$ and $\partial_zA_0(z_h)$ can be tuned to obtain a correct behavior near $z=\varepsilon$. In our calculation, we fix the charge density $\rho=1$. The coupled differential equations allow multiple solutions. The solution with lower energy configuration is the one with monotonically increasing $\psi$ from UV boundary to horizon. Other solutions with a greater number of nodes are for higher energy configurations, so thermodynamically unfavorable. The Cooper pair condensate can be obtained by
	\begin{align}
		\left<O_2\right>=\sqrt{2}\psi^{(2)}\,,
	\end{align}
	where we follow the normalization in~\cite{Hartnoll:2008kx}. Numerical calculations for various temperature show that $\left<O_2\right>=0$ when $T$ greater than a critical value, $T>T_c$ and $\left<O_2\right>>0$ when $T<T_c$.
	
	We can calculate conductivity in the presence of the oscillating electric field, by assuming the following solution
	\begin{align}
		A_x(t,z)=e^{-i\omega t}A(z)\,,
	\end{align}
	while $A_y=A_z=0$. This allows as to calculate only the conductivity in $x$ direction, without losing any generality because of symmetry. The equation of motion for the gauge field is given by
	\begin{align}
		\partial_z^2A+\frac{f'}{f}\partial_z A+\left(\frac{\omega^2}{f^2}-2\frac{\psi^2}{z^2f}\right)A=0\,. \label{Aeom}
	\end{align}
	In the absence of Cooper pair condensate, $\psi=0$, one obtains two possible solutions~\cite{Albrecht:2012ek}, which correspond to wave going into the horizon and wave going away from the horizon. However, in the outgoing-wave solution unusual situation occurs where the direction of the resulting current is opposite to the direction of the electric field. This provides a motivation to choose the ingoing-wave boundary condition for the bulk equation of motion above. The gauge field near horizon should satisfy the following equation for the ingoing-wave
	\begin{align}
		A'(z\rightarrow z_h)=\frac{i\omega}{f}A\,,
	\end{align} 
	which is consistent with (\ref{Aeom}) when $\psi=0$. 
	
	Near the UV boundary, the gauge field behaves as
	\begin{align}
		A_x=A^{(0)}_x+A^{(1)}_x z+\ldots\,,
	\end{align}
	where $A^{(0)}_x$ is the vector potential of the electromagnetic field and $A^{(1)}_x=\left<J_x\right>$ is the expectation value of the current. The electric field can be obtained by
	\begin{align}
		E_x=-\dot{A}^{(0)}_x=i\omega A^{(0)}_x\,.
	\end{align}
	Hence, one obtains the conductivity
	\begin{align}
		\sigma=\frac{J_x}{E_x}=-i\frac{A^{(1)}_x}{\omega A^{(0)}_x}\,.
	\end{align}
	\section{Vortex solution}
	For vortex solutions, we use coordinates in the form of $(t,z,r,\phi)$, where $r$ and $\phi$ are the radial and tangential direction in the 2D plane, respectively. 
	Solutions that preserves global U$(1)$ and rotation symmetry in the 2D plane should be in the following form~\cite{Montull:2009fe}
	\begin{align}
		\psi=\psi(r,z) e^{in\phi}\,,\quad A_0=A_0(r,z)\,,\quad A_\phi=A_\phi(r,z)\,,
	\end{align}
	and all other fields are zero. Integer $n$ is the winding number which determines different topological solutions.
	
	With the above ansatz, we obtain the following equations
	\begin{align}
		\psi_{zz}+\frac{z^2}{f}\partial_z\left(\frac{f}{z^2}\right)\psi_z+\frac{1}{f}\psi_{rr}+\frac{1}{fr}\psi_r+\left(\frac{A_0^2}{f^2}-\frac{(A_\phi-n)^2}{r^2f}+\frac{2}{z^2f}\right)\psi&=0\,,\\
		{A_{0}}_{zz}+\frac{1}{f}{A_0}_{rr}+\frac{1}{rf}{A_0}_{r}-\frac{2}{z^2f}A_0\left|\psi\right|^2&=0\,,\\
		{A_\phi}_{zz}+\frac{\partial_zf}{f}{A_\phi}_{z}+\frac{1}{f}{A_\phi}_{rr}-\frac{1}{rf}{A_\phi}_r-\frac{2\left|\psi\right|^2}{z^2f}(A_\phi-n)&=0\,.
	\end{align}
	For our choice of $m^2=-2$ in equation (\ref{matteraction}), the UV boundary conditions are~\cite{Montull:2009fe}
	\begin{align}
	\partial_z\psi\big|_{z=0}=0\,,\quad \partial_zA_0\big|_{z=0}=-\rho\,,\quad A_{\phi}\big|_{z=0}=\frac{1}{2}r^2 B\,.\label{boundarycond}
	\end{align} 
	where $B$ is the external magnetic field. For a regular solution we impose at the horizon $A_0|_{z=z_h}=0$ and at the $r=0$, we require for $n\neq0$
	\begin{align}
	\psi\big|_{r=0}=0\,,\quad \partial_rA_0\big|_{r=0}=0\,,\quad A_{\phi}\big|_{r=0}=0\,,
	\end{align}  
	while for $n=0$, $\partial_r\psi|_{r=0}=0$. We consider a superconductor material that extend to $r=R$ where $R$ is much bigger than the vortex size. Hence, we will set $\rho=1$ when $r<R$ and  $\rho=0$ for $r>R$.
	
	\section{Deconstruction}
	
	Discretization of the action (\ref{matteraction}) is given by
	\begin{align}
		S=\sum_{j=1}^{N-1}a_j\int d^3x& \bigg\{\frac{1}{2}\left(\partial_0 A_{zj}-\partial_zA_0\right)^2+\frac{1}{2f_j}(F_{0a})_j^2-\frac{f_j}{2}\left(\partial_a A_{zj}-\partial_zA_{aj}\right)^2\nonumber\\
		&-\frac{1}{4}\left(F_{ab}\right)^2_j+\frac{1}{z_j^2f_j}\left|\left(\partial_0-iA_{0j}\right)\psi_j\right|^2-\frac{f_j}{z_j^2}\left|\left(\partial_z-iA_{zj}\right)\psi_j\right|^2\nonumber\\
		&-\frac{1}{z_j^2}\left|\left(\partial_a-iA_{aj}\right)\psi_j\right|^2-\frac{m^2}{z_j^4}\left|\psi_j\right|^2
		\bigg\}\,.\label{actiondescretized}
	\end{align}
	In the action above, the extra-dimension has been replaced with lattice of $N$ points
	\begin{equation}
		z_j=\bigg\{
		\begin{array}{ll}
		\varepsilon+(j-1)a,&{\rm for}\, j=1\ldots N-1\\
		\varepsilon+(N-2)a+a_h,&{\rm for}\,j=N\,,
		\end{array}
	\end{equation}
	where $z_N=z_h$, $f_j=f(z_j)$. In the discretized action, derivative is defined as
	\begin{align}
		\partial_z\psi_j=(\psi_{j+1}-\psi_j)/a_j\,.
	\end{align}
	
\begin{figure}
	 	\centering
	 	\includegraphics[width=10 cm]{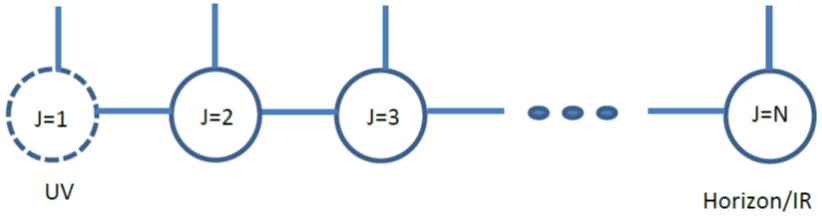}
	 	\caption{Diagram representation of the deconstructed theory.}
	 	\label{moosediagram}
	 \end{figure}

	As in~\cite{Albrecht:2012ek}, we consider a deconstructed theory that will reproduce the discretized action. Consider 3D theory with $N-1$ gauged U$(1)$ symmetry and one global U$(1)$ symmetry as represented in Figure \ref{moosediagram}. The leftmost dashed-circle represents the global U$(1)$ symmetry in the UV boundary, while the remaining solid circles which represent the bulk lattice points are labeled with $j=2\ldots N$. They are gauged U$(1)$ symmetry. Horizontal lines connecting the circles represent a set of bi-fundamental complex scalar fields $\Sigma_j, j=1\ldots N-1$, which transforms under the $j$-th and $(j+1)$-th U$(1)$ factors with charges $-1$ and $+1$ respectively. In addition, there are $N$ complex scalar fields, $\psi_j, j=1\ldots N$, which transforms under the $j$-th U$(1)$ factor, with charge $+1$.
	
	The Lagrangian for the 3D theory of the deconstructed model is the following~\cite{Albrecht:2012ek}
	\begin{align}
		\mathcal{L}=\sum_{j=2}^{N-1}\left[-\frac{1}{4}(F_{\mu\nu})_j(F^{\mu\nu})_j+Z_j \left|D_\mu\psi_j\right|^2\right]+\sum_{j=1}^{N-1}\left[\left|D_\mu\Sigma_j\right|^2-Z_jV_j\right]\,.\label{actiondeconstruct}
	\end{align}
	Contraction of the Greek index uses discretized ``metric'' $g^{\mu\nu}_j$ which should be considered just as  Lorentz violating couplings of the theory. Constant $Z_j$ and potential $V_j$ are chosen to reproduce continuum limit. One obtains,
	\begin{align}
		g^{00}_j=\frac{1}{f_j}\,,\quad g^{ab}_j=-\delta^{ab}\,,\quad Z_j=\frac{1}{z^2_j}\,.
	\end{align} 
	The potential is chosen to be,
	\begin{align}
		V_j=\begin{array}{ll}
		2|v_j\psi_{j+1}-\psi_j\Sigma_j|^2\,,&{\rm for}\quad j=1\\
		2|v_j\psi_{j+1}-\psi_j\Sigma_j|^2+m^2|\psi_j|^2/z^2_j\,,&{\rm for}\quad j=2\ldots N-1\,,
		\end{array}
	\end{align}
	where $v_j=\frac{1}{a_j}\sqrt{\frac{f_j}{2}}$. The bi-fundamental scalar fields can be written as
	\begin{align}
		\Sigma_j=v_j\exp[ia_jA_{zj}]\,,
	\end{align}
	with vacuum expectation value $\left<\Sigma_j\right>=v_j$.
	When we set $\Sigma_j$ equal to vev in (\ref{actiondeconstruct}), the discretized action (\ref{actiondescretized}) is reproduced. 
	
	The equation of motion for the deconstructed model can be written as
	\begin{align}
	\psi_j^{zz}+\frac{1}{a_j}\left(1-\frac{z_j^2f_{j-1}}{z_{j-1}^2f_j}\right)\psi_{j-1}^z+\frac{1}{f_j}\psi_j^{rr}+\frac{1}{rf_j}\psi_j^r+\left(\frac{A_{0,j}^2}{f_j^2}-\frac{(A_{\phi,j}-n)^2}{f_j r_j}+\frac{2}{z_j^2f_j}\right)\psi_j&=0\,, \label{deconstructeom}\\
	A^{zz}_{0,j}+\frac{1}{f_j}A^{rr}_{0,j}+\frac{1}{rf_j}A^r_{0,j}-\frac{2}{z^2f_j}A_{0,j}\left|\psi_j\right|^2&=0\,, \label{deconstructeom2}\\
	A^{zz}_{\phi,j}+\frac{1}{a_j}\left(1-\frac{f_{j-1}}{f_j}\right)A^z_{\phi,j}+\frac{1}{r}A^{rr}_{\phi,j}-\frac{1}{rf_j}A_{\phi,r}-\frac{2\psi_j^2}{z^2f_j}\left(A_{\phi,j}-n\right)&=0\,, \label{deconstructeom3}
	\end{align}
	where partial derivative for the $r$ coordinate is written in the form of $\psi^{r}_j\equiv \partial_r\psi_j$ and $\psi^{rr}_j\equiv \partial_r^2\psi_j$. We also define  $\psi^z_j=(\psi_{j+1}-\psi_j)/a_j$, $\psi^{zz}_j=(\psi_{j+1}-2\psi_j+\psi_{j-1})/a^2$, for $j=2\ldots N-2$ and $\psi^{zz}_{N-1}=((\psi_N-\psi_{N-1})/a_h-(\psi_{N-1}-\psi_{N-2})/a)/a_h$. Similarly for other fields. In addition to the above equations, we have the following equations near horizon 
	\begin{align}
	-\frac{3}{z_h}\psi^z_{N-1}+\psi^{rr}_N+\frac{1}{r}\psi^r_N-\left(\frac{(A_{\phi,N}-n)^2}{r^2}-\frac{2}{z_h^2}\right)&=0\label{deconstructeomboundary}\\
	A^{zz}_{0,N-1}+\frac{z_h}{3}\left(A^{rr}_{0,N-1}+\frac{1}{r}A^r_{0,N-1}\right)-\frac{\psi_N^2}{3z_h}A_{0,N-1}&=0\label{deconstructeomboundary2}\\
	-\frac{3}{z_h}A^z_{\phi,N-1}+A^{rr}_{\phi,N}-\frac{1}{r}A^r_{\phi,N}&=0\label{deconstructeomboundary3}
	\end{align}
	
	The discretized boundary conditions near horizon are 
	\begin{align}
	A_{0,N}=0\,,\qquad \psi_{N-1}=\left(1-\frac{2a_h}{3z_h}\right)\psi_N\,,
	\end{align}
	and near UV boundary the conditions are
	\begin{align}
	\qquad A_{0,2}-A_{0,1}=-a_1\rho\,,\qquad \psi_2=\psi_1\frac{(\varepsilon+a)^2}{\varepsilon^2}
	\end{align}

	\section{Numerical Results}

\begin{figure}
		\centering
		\includegraphics[width=9 cm]{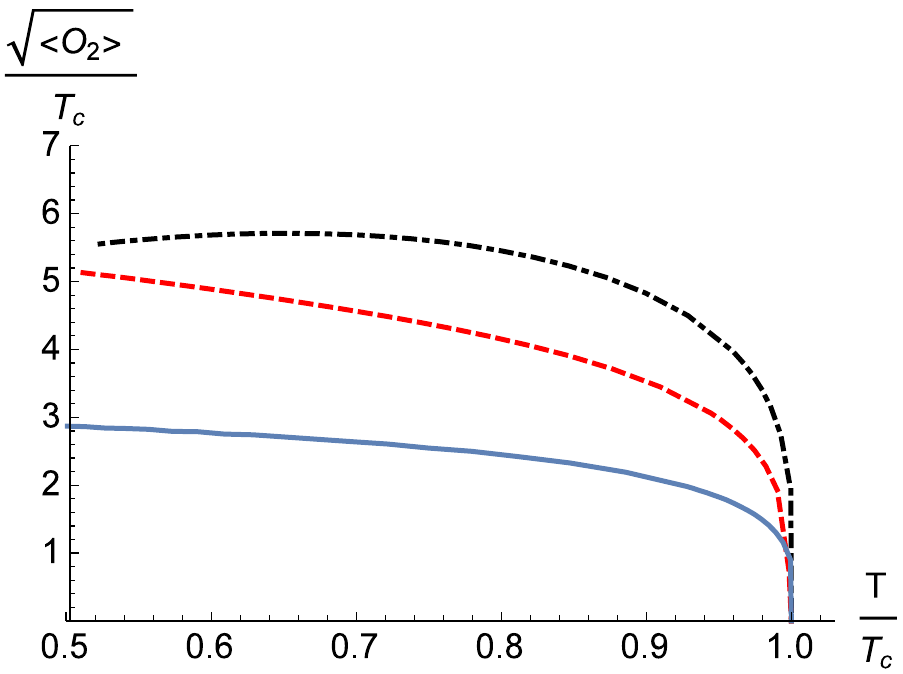}
		\caption{Condensate as a function of temperature $T$ when we turn off the external magnetic field $B=0$.  From top to bottom, the dot-dashed line corresponds to $N=5$, the dashed line for $N=10$, and the solid line for $N=100$.}
		\label{condensatedeconstruct}
	\end{figure}

	First we consider no external magnetic field, $B=0$. To implement this, we set $A_{\phi,j}=0$ and $n=0$ in equation (\ref{deconstructeom}) to equation (\ref{deconstructeomboundary3}). In Figure \ref{condensatedeconstruct} we show vacuum expectation value of the Cooper pair operator as a function of the temperature for $N=5$, $N=10$ and $N=100$ lattice points. The expectation value of the operator vanish for temperature above  certain critical values, $T_c$. For our numerical calculations, we set the UV cutoff $\varepsilon=a$. We also fix the lattice spacing at the horizon $a_h=10^{-5}$. We find $T_c=0.083\sqrt{\rho}$, $T_c=0.097 \sqrt{\rho}$, $T_c=0.122\sqrt{\rho}$, for $N=5$, $N=10$ and $N=100$, respectively.
	
	In the case of non-zero external magnetic field with vortex solution, for our numerical calculation we choose~\cite{Montull:2009fe}
	\begin{align}
		R=\frac{50}{\sqrt{\rho}}\,, \quad T=0.065\sqrt{\rho}\,.
	\end{align}
	Total flux of the magnetic field going through the vortex is quantized as $2\pi n$, hence the magnetic field is given by
	\begin{align}
	B_n=\frac{2n}{R^2}\,.
	\end{align}
	  For our numerical calculations, the $r$ direction is discretized with $N_r=800$ lattice points. We numerically solve the coupled non-linear equations (\ref{deconstructeom}) and (\ref{deconstructeomboundary}) using relaxation methods. The calculation steps are as follows. First we pick a reasonable guess function, then each lattice point is updated starting from the horizon ($j=N$, $r=0$) to ($j=N$, $r=R$) . Then continue to ($j=N-1$, $r=0$) until ($j=N-1$, $r=R$) and so on until the whole lattice points covered. The process is repeated until the fields are stabilized.  
	
\begin{figure}
	 	\centering
	 	\includegraphics[width=9 cm]{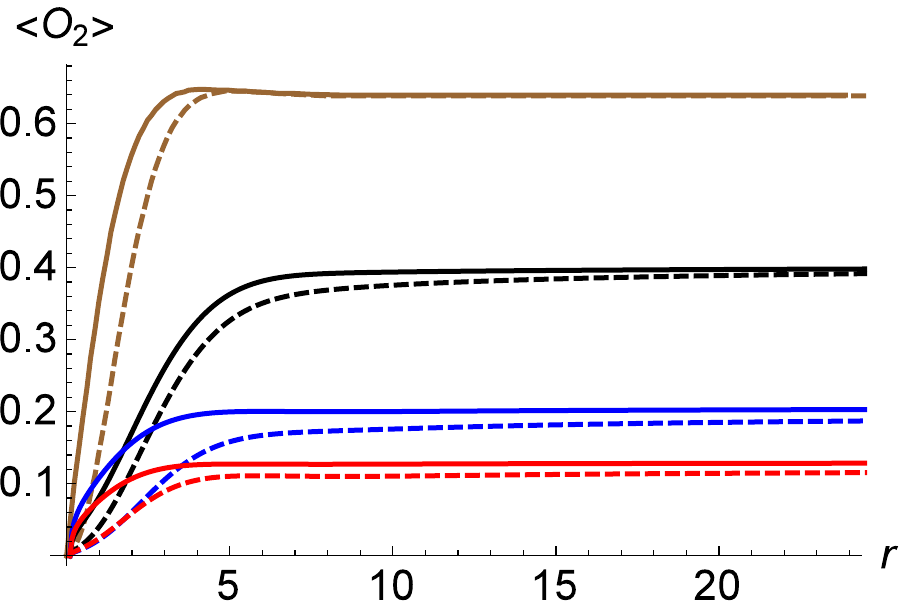}
	 	\caption{Condensate as a function of $r$. The solid lines are for the winding number $n=0$ while the dashed lines are for the $n=1$. From top to bottom for large $r$ correspond to the continuum limit, $N=5$, $N=10$, and $N=100$.}
	 	\label{condensatevortex}
	 \end{figure}

	In Figure \ref{condensatevortex} we show the vacuum expectation value of the Cooper pair operator for $N=5, 10$ and $100$ lattice points. 
	
	\section{Conclusions}
	
	In this paper, we have followed the AdS/CFT prescription for lower dimensional discretized theory given in~\cite{Albrecht:2012ek}. We calculate the Cooper pair condensate for the vortex solution in a deconstructed holographic model of superconductor for $N=5$, $N=10$ and $N=100$. We have shown that even with a small number of lattice points, the model reproduces enough physics of the corresponding continuum holographic model.

	This work was supported by Fundamental Science Grant, Department of Research, Technology and Higher Education of Indonesia, No. DIPA-023.04.1.673453/2015.
	
\bibliographystyle{JHEP}

\bibliography{superconductingvortex}
	
\end{document}